\documentstyle[twocolumn,fleqn,ndes,psfig]{article}

\title{Chaotic dynamics and coherent structure in electron beam with
virtual cathode in the diode with local neutralization}
\author{V. G. Anfinogentov \and
   A. E. Hramov}
\affiliation{
College of Applied Science, Saratov State University,\\
Astrakhanskaya 83, Saratov
410026, RUSSIA\\ 
aeh@cas.ssu.runnet.ru}

\begin{document}

\maketitle

\begin{abstract}
The paper consider a complex dynamics of electron beam with virtual
cathode and local neutralization of the beam charge density near anode.
Different types of nonlinear behaviour, including deterministic chaos,
were treated. It is shown that chaotic dynamics arises as results of
spatiotemporal structures interaction. 
\end{abstract}

\subsection{Introduction and model}
In the present paper  we consider a electrostatic short-circuited
diode model. The immobile ion background with concentration $n_p$
locates near left boundary (anode plasma). The electron beam with
overcritical current injected into diode with nonpurtubated velocity
$v_0$ and charge density $\rho_0$. In this case virtual cathode (VC)
forms in the beam as a result of the electrostatic instability \cite{r1}, and
some beam part is reflected from VC to injected boundary. VC oscillates
in the diode region.
The electron beam with VC demonstrates wide diversity of nonlinear 
phenomenons, including chaotic behaviour \cite{r2}, synchronization 
\cite{r3} and other. Investigation of complex dynamics attracts many
researchers, that is, such behaviour is the characteristic property of
the beam with VC. That counts is study the structure formation, since
it is well known, that chaotic dynamics in distributed systems is connected
with pattern formation (see \cite{r4} and references therein).

The such simplest model of device with VC as a planar diode with
overcritical current describes a different nonlinear phenomenons  in
the real vircator systems. Our model with local neutralisation is
a simple model of vircator with injected plasma \cite{r2}.

The behaviour of system is determined by the dimensionless parameter
related current
\[
\alpha = \omega_p L/v_0,
\] 
where $\omega_p$ is the beam
plasma frequency, $L$ is the distance between diode planes and 
neutralisation parameter 
\[
n=n_p/n_0.
\] 
Hence $n_0=\rho_0/e$ and value
of plasma region length $x_p$ is constant ($x_p = 0.25L$).

The effect of neutralisation degree of anode plasma on VC
dynamics was investigated  with the aid of {\it particle--in--cell}
simulation. The macroparticles in the simulation obey the non-relativistic
equations of motion
\[
\begin{array}{lcl}
dx/{dt}&=&v,\\
{dv}/{dt}&=& -({q}/{m}){\partial \phi}/{\partial x},\\
\end{array}
\]
where $x$ is the position of particles, $v$ is the velocity of of the
particles, $q$ is the charge and $m$ is the mass of the macroparticles. The
code integrates the equation of motion forward in time using a leapfrog
scheme. The potential $\phi$ is computed by the Poisson's equation in one
dimension
\[
\frac{\partial^2 \phi(x)}{\partial x} = -\alpha^2(\rho(x)-\rho_p(x)).
\]
Hence $\rho(x)$ is the spatial distribution of beam charge density and
$\rho_p(x)$ is the distribution of immobile ion background.
In our case
\[
\rho_p(x)=\left\{
\begin{array}{ll}
n \cdot e, & x \le x_p, \\
0,         & x > x_p.   \\
\end{array}
\right.
\]
\vskip4mm

\subsection{System dynamics}

The tentative analysis of nonlinear dynamics were effectuated
from observation  of time series of electric field oscillation in the
injection plane. Power spectra and projections of attractors were
reconstructed from time series. Based on this analysis, domains for distinct
behaviour were isolated in parameter plane ($\alpha$, $n$) (see Fig.~1).

\begin{figure}[t]
\vskip 0mm
\centerline{\psfig{figure=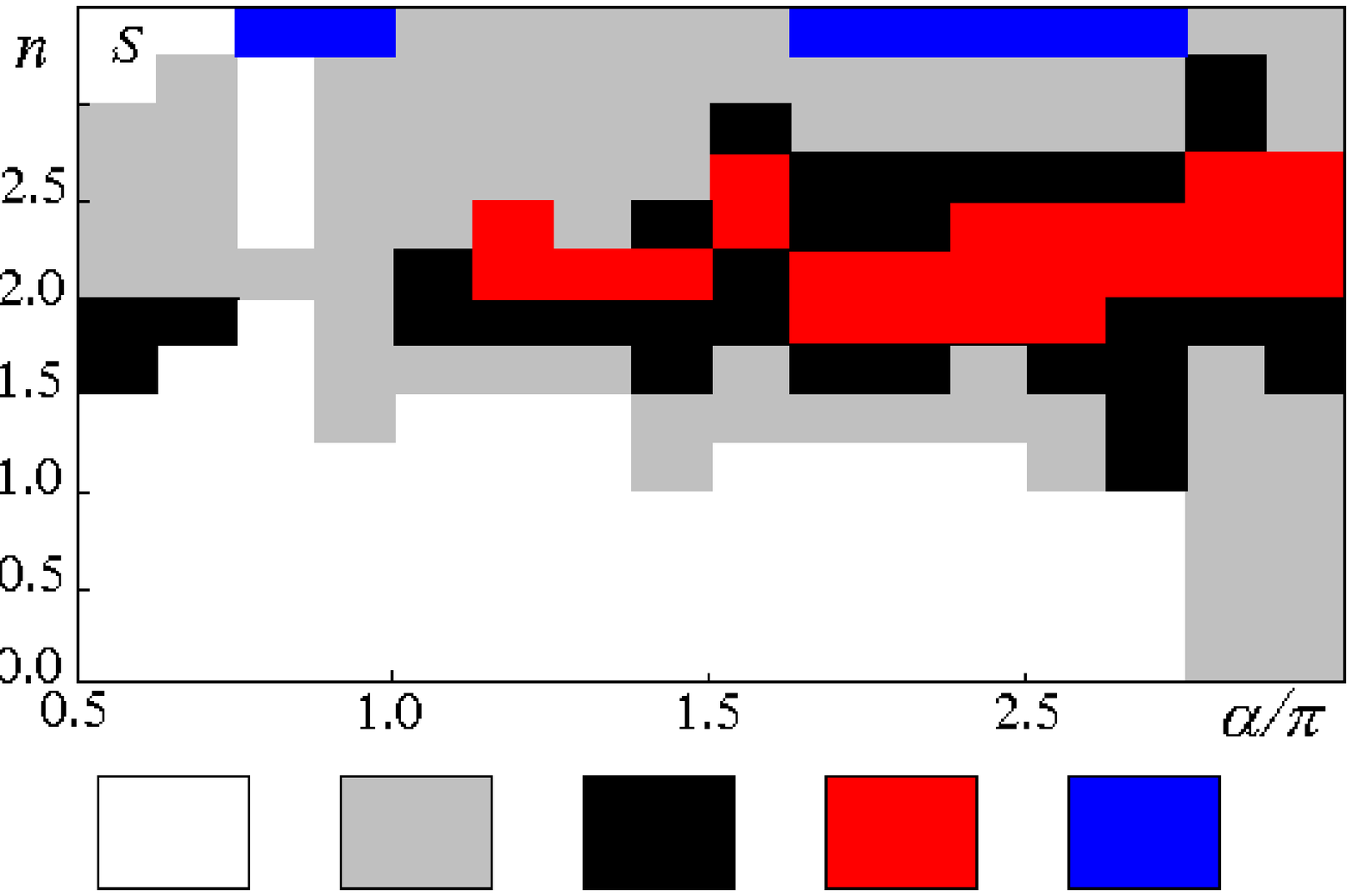,height=45mm}}
\vskip -1 mm
\centerline{\it  \hskip-1mm A \hskip9mm  B 
                  \hskip9mm C \hskip9mm D \hskip9mm E}
\vskip 3mm
\noindent
{\bf Fig.~1:} Bifurcation diagram on the parameter space ($\alpha$, $n$). White
area {\it S} corresponds to nonuniform equilibrium
\vskip3mm
\end{figure}

\begin{figure}[t]
\centerline{\psfig{figure=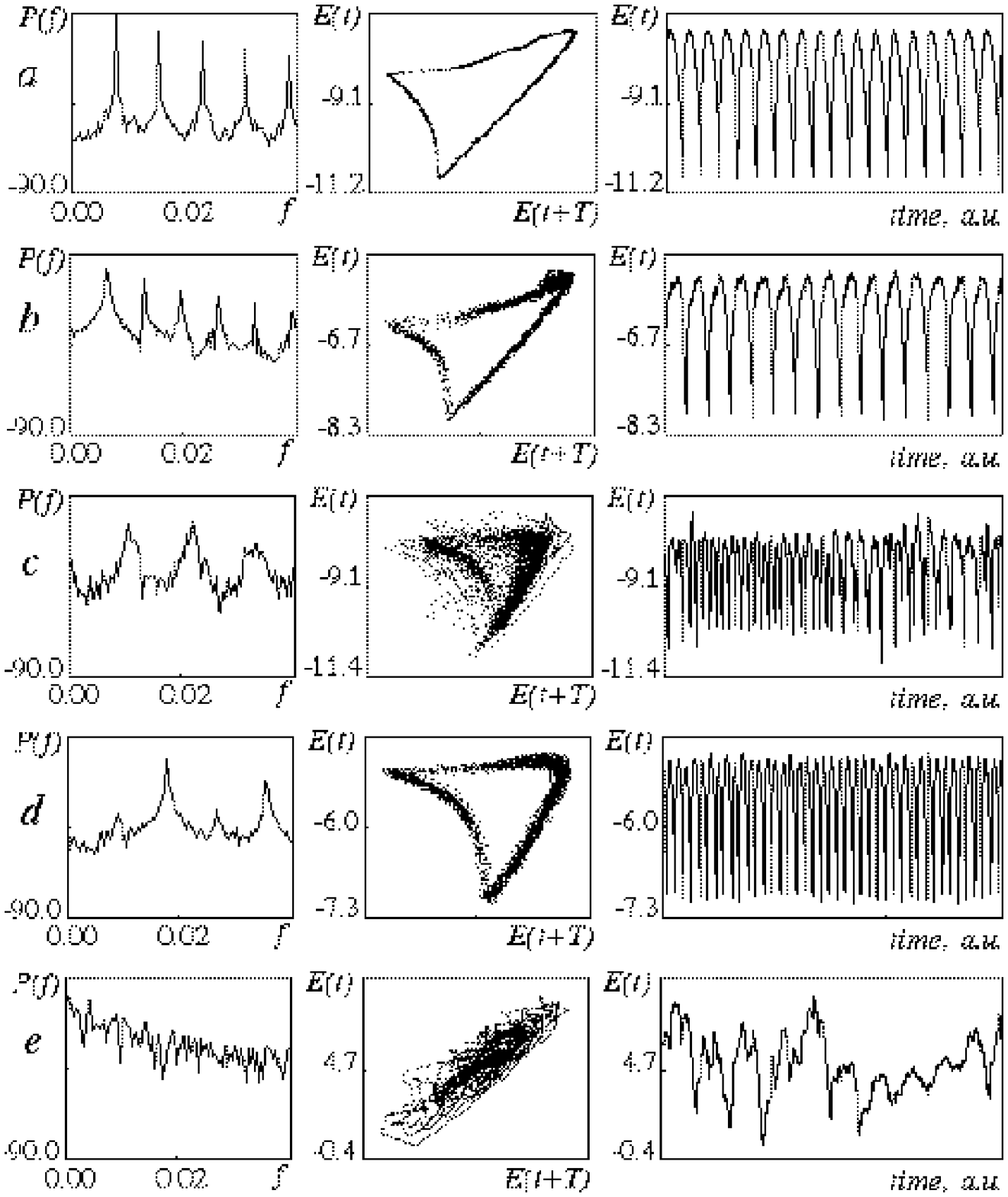,height=95mm}}
\vskip 3mm
\noindent
{\bf Fig.~2:} {Power spectra, reconstructed phase portraits and
time series for different regimes}
 
\vskip3mm
\end{figure}

The VC oscillation (VCO) for small value of neutralisation is regular
(domain marked {\it A} in the parameter plane; Fig.~2(a)). 
Analysis of physical processes shows that only one electron bunch (VC) 
is arisen
in the system. This bunch is marked on the spatiotemporal diagram
(Fig.3(a)). Besides, metastable particles, which  exist in the interaction
space during of more than one period of VCO, is observed in the beam.
However, charge density of the metastable bunch is small, and it is little
affected by VCO. The weakly chaotic VCO arises as neutralisation and current
increase (domain {\it B} for $n<2.0$; Fig.~2(b)). In this case metastable
bunch density grows. The further increasing of $n$ leads to formation of
profound metastable bunch in the beam (Fig.~3(b)). A buildup of space charge
density in VC region entails the regime with large base frequency in the VCO
spectrum (compare Fig.~2(a,b) and Fig.~2(d), that obtain for the same value
of $\alpha=2.125\pi$). This behaviour of system take place for $n>2.25\div2.5$ ({\it
B}; Fig.~2(d)). A change-over from the weak chaos for small values of $n$ to
the weak chaos for large $n$ derives through two domains of strongly chaotic
VCO. In the first regime (domain {\it C} in Fig.~1) phase portrait is homogeneous, there are
not sharp peaks in the power spectrum. The second regime ({\it D}) may be
treated as intermittence (Fig.~2(c)).

For large values of $n$ (domain {\it E}) system demonstrates highly
non-regular oscillation with noise-like spectrum and homogeneous attractors (Fig.~2(e)). 
In this case VC is formed out of anode plasma region. 
VC exists constantly and chaotic dynamics is determined by the reflection of
the particles from VC. Note, VC is not moved in the space, but depth of
potential barrier is oscillated in time.

Dimensions of the reconstructed attractors was estimated for different types
of chaotic behaviour. Fig.~4 presents correlation dimension of attractors
$D$ {\it versus} value of embedded dimension $m$ for strong chaos. Dimension
is saturated for small values of $m$. Small values of embedded dimensions
justify appearance of chaotic behaviour in the beam in the result of
interaction between small numbers of structures.

\vskip -3mm
\centerline{\psfig{figure=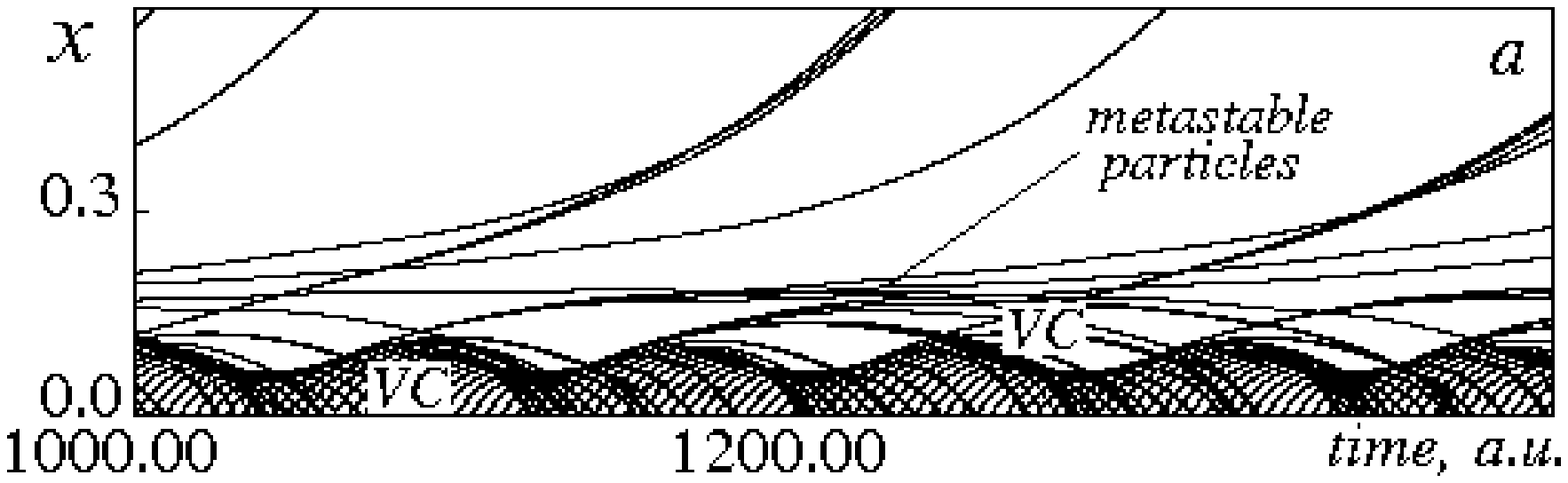,height=23mm}}
\centerline{\psfig{figure=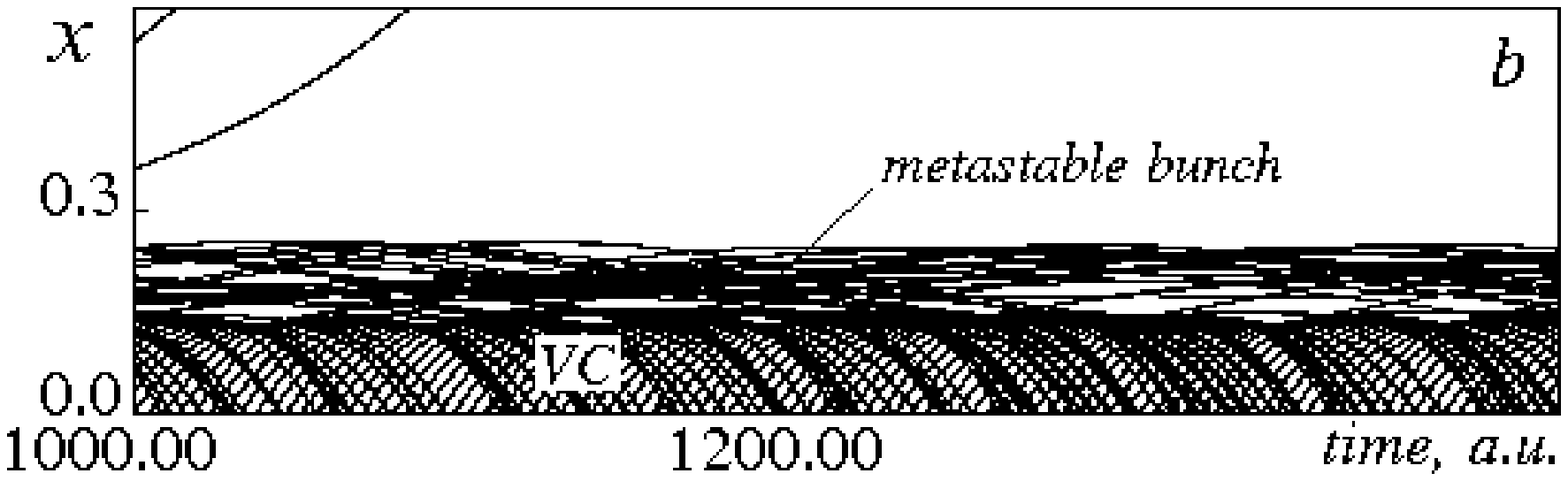,height=23mm}}
\vskip -1 mm
\vskip 3mm
\noindent
{\bf Fig.~3:} {Spacetime diagrams for regular (a) and chaotic (b) oscillation}
\vskip3mm

\vskip 4mm
\centerline{\psfig{figure=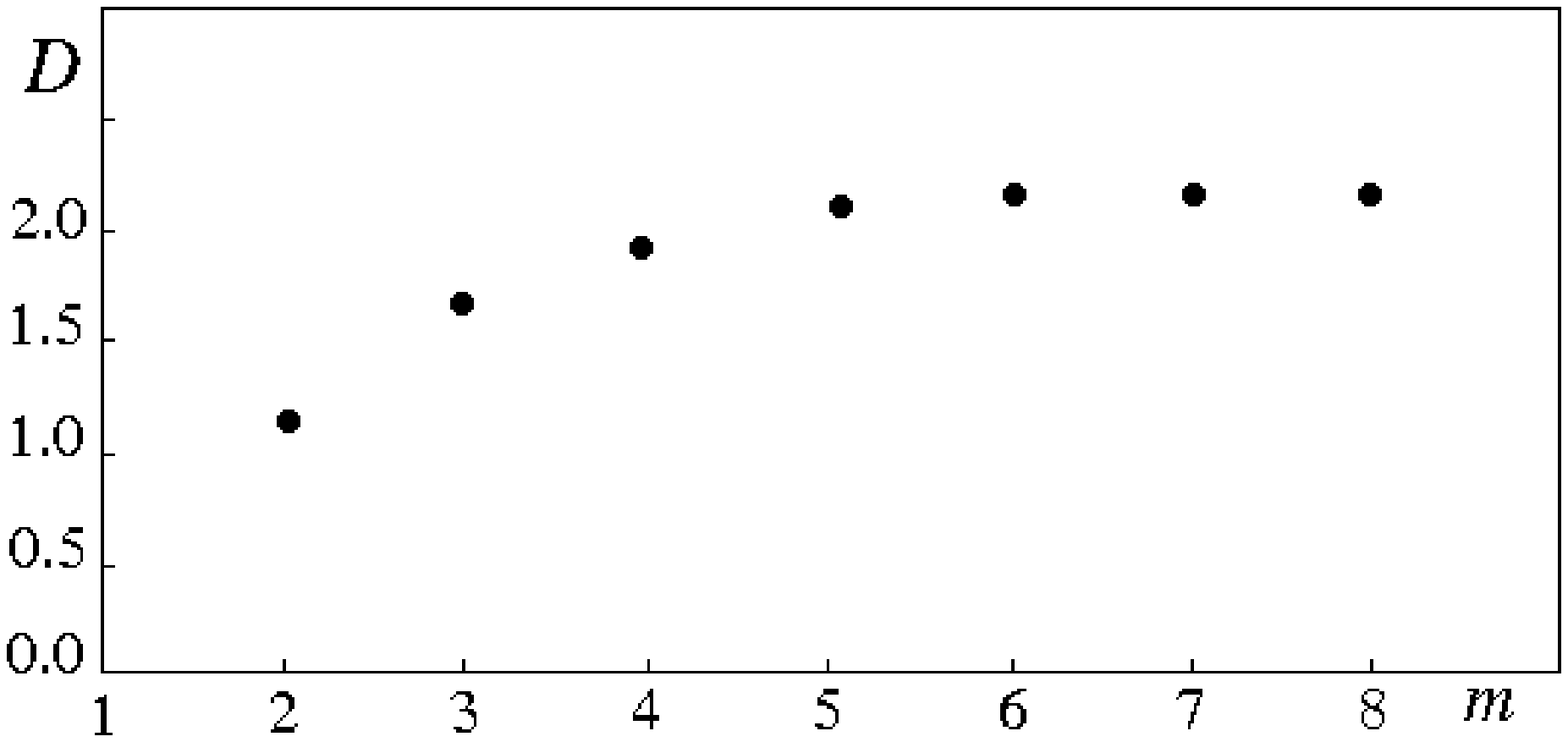,height=35mm}}
\vskip -1 mm
\vskip 3mm
\noindent
{\bf Fig.~4:} {Dependence of attractor dimension from embedded dimension}
\vskip3mm

\subsection{Pattern formation}

The spatiotemporal data of charge density $\rho(x,t)$ were analyzed by
the Karhunen--Loeve orthogonal decomposition \cite{r5}.
This method decomposes a data set into spatial orthogonal modes
$\{ \psi(x)\}_i$. This modes is the solution of the integral equation
\[
\int R(x,x')\psi(x')dx' = \Lambda \psi(x),
\]
where $R(x,x') = \langle \rho(x,t) \rho(x',t) \rangle_t$
is the mutual correlation function.
The value of eigenvalue $\Lambda_i$ is proportional to the energy
of $i$-th mode. Karhunen--Loeve method is optimal in the sense that one is
optimized (from the viewpoint of energy of modes) eigensets
$\{ \Lambda\}_i$ and $\{ \psi\}_i$. The measure of energy of modes is
\[
W_i = \frac{\Lambda_i}{\sum_k \Lambda_k}.
\]

The energy of several first modes for different values
of $n$ is presented  in Table for $\alpha=1.35\pi$. The typical spatial distributions of
modes are shown in Fig.~5. In the all regimes of VCO behaviour of beams is
determined by the small number of structures, because more then 90\% energy
is contained in the $3\div4$ higher modes.

\begin{table}[h]
\vskip 3mm
\noindent
{\bf Table:} Energy $W_i$ (in \%) of Karhunen--Loeve modes
\vskip 3mm
\begin{tabular}{|c|ccccc}
\hline
\multicolumn{1}{|c|}{Number of}& \multicolumn{5}{c}{Value of neutralisation, $n$}\\
\cline{2-6}
mode, $i$ & 0.0 &0.35&0.7&1.05&1.4 \\
\hline 
1 &  67.1 & 62.1& 60.0& 56.1 & 52.6\\ 
2 &  17.1 & 16.4& 20.0& 21.8 & 23.3 \\ 
3 &  5.7 & 8.3& 7.1& 7.9 & 8.5       \\ 
4 &  3.1 & 4.6& 4.1& 4.6 & 5.0        \\ 
\hline
\end{tabular}

\begin{flushright}
\begin{tabular}{cccccc|}
\hline
 \multicolumn{6}{c|}{Value of neutralisation, $n$}\\
\hline
1.75 &2.1 &2.45 &2.8 &3.15 &3.5 \\
\hline 
52.3 &60.8 &71.5 & 70.3 & 69.8 &
83.2 \\ 
25.1 &21.4 &10.2 & 12.3 & 12.7 &
11.5 \\  
 8.5 & 7.4 &6.3 & 4.2 & 4.1 &
3.0 \\ 
 4.6 &2.9 &2.2 & 3.3 & 3.6 &
0.4 \\ 
\hline
\end{tabular}
\end{flushright}
\end{table}

For small neutralisation first modes demonstrates strong nonuniform charge
density distribution with one peak (Fig.~5(a)), that corresponds of typical distribution
of density in VC. Second mode describes  are processes of destruction of VC
and disposal of particles from VC to anode. Third and other modes
correspond to additional bunches in the beam. Increasing of $n$  leads to
growth of second modes at the expense of higher mode.

\begin{figure}[htb]
\vskip -1mm
\centerline{\psfig{figure=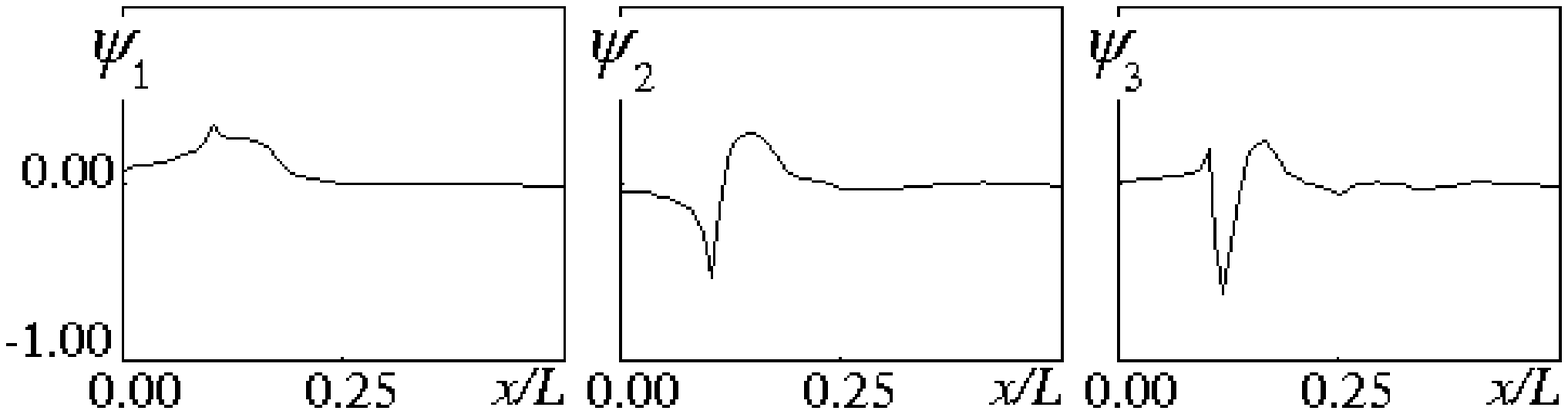,height=20mm}}
(a) $n=0.25$ \\ \vskip2mm
\centerline{\psfig{figure=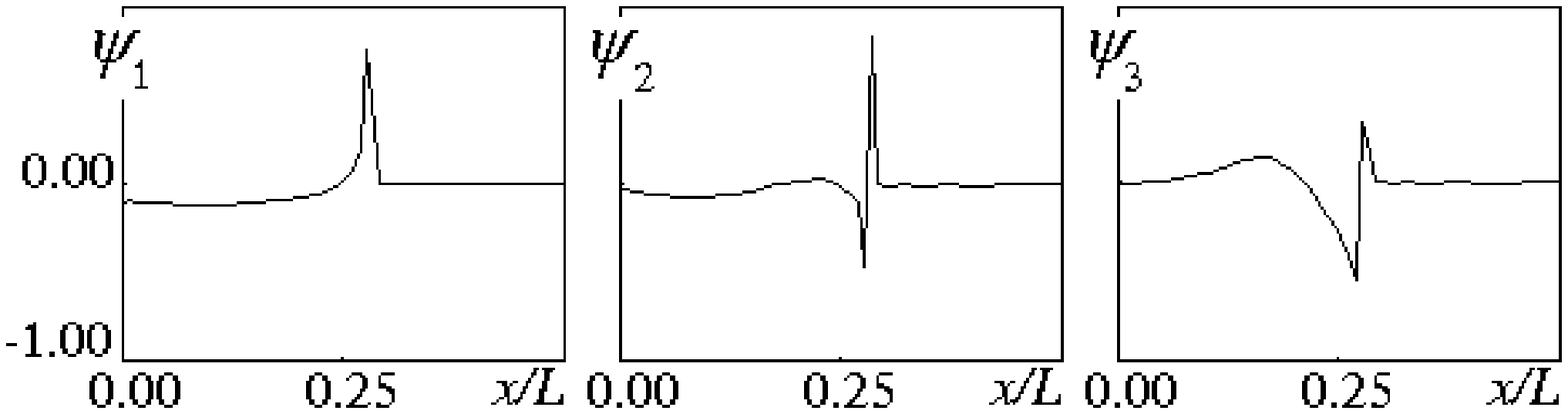,height=20mm}}
(b) $n=2.5$ \\ \vskip2mm
\centerline{\psfig{figure=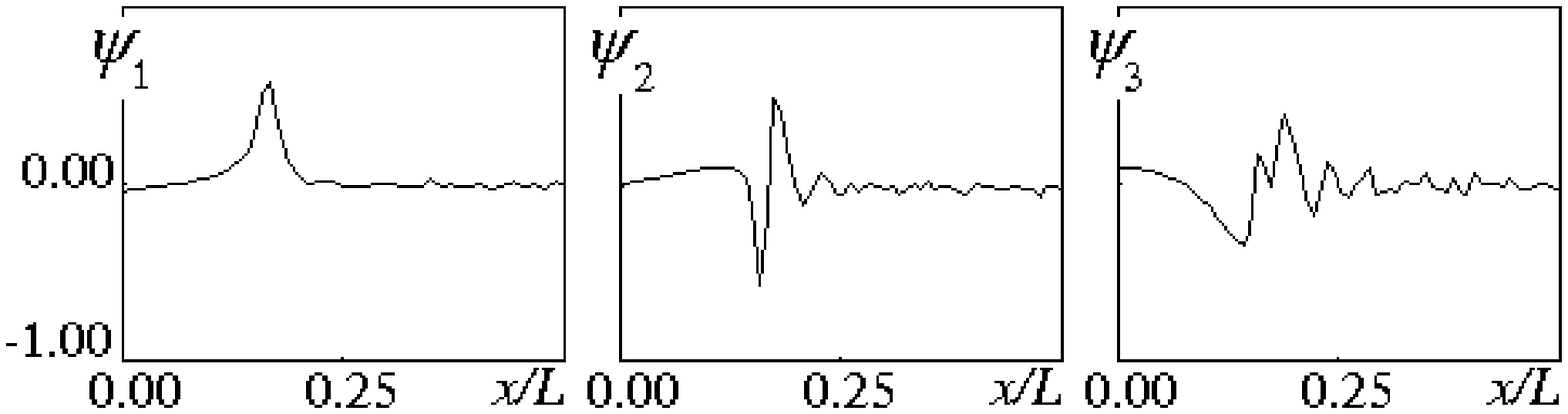,height=20mm}}
(c) $n=3.25$ \\
\vskip -1mm

\vskip 3mm
\noindent
{\bf Fig.~5:} {Typical structures for different values of neutralisation}
\vskip3mm

\end{figure}

Metastable bunch is formed for values of $n\sim 2\div3$. In this case first
and second modes together describes dynamics of VC and metastable bunch,
besides cross-correlation between  dynamics modes $A_1(t)$ and $A_2(t)$ is
large. Hence temporal dynamics of modes is 
\[
A_i(t)=\int \rho(x,t)\psi_i(x)dx.
\]

Spatial distribution of modes is strongly localizing in the interaction space
for regime {\it E} (Fig.~5(c)). Only one structure --- weakly oscillating in the space virtual
cathode --- exists in the beam, and more than 80\% energy accumulated in the
first mode. Chaotic VCO is determined by the reflection of the larger part
of beam from VC.

\subsection{Conclusions}

In the electron beam with VC and local neutralisation different types of
nonlinear oscillation are recognized. Influence of density of 
anode plasma on the chaotic dynamics of VC are considered. Large
neutralisation degree leads to strong chaos in the VCO. Strange 
attractor is most homogeneous in this case. Relationship between 
different type of chaotic behaviour and structure formation are shown 
with the help of analysis of physical processes in the diode.
The typical patterns  (VC and different additional structures) 
were recognized for regimes with small and 
large neutralisation. It is shown that appearance of chaotic 
behaviour connects with growth of charge density in the additional 
structures.
Investigation of chaotic behaviour in our model may be helpful for 
design of vircators with anode plasma grid, because changing of 
density of anode plasma lead to changing of radiation characteristics in 
these devices.

\subsection{Acknowledgment} 
This work was supported by Russian Foundation of Fundamental Research (Grant
No~96-02-16753).

\end{document}